# Cognitive Forwarding Control in Wireless Ad-Hoc Networks with Slow Fading Channels

Vesal Hakami and Mehdi Dehghan[*]

Department of Computer Engineering and Information Technology, Amirkabir University of Technology

424 Hafez Avenue, Tehran, Iran

{vhakami, dehghan}@aut.ac.ir

**Abstract**

We propose a decentralized stochastic control solution for the broadcast message dissemination problem in wireless ad-hoc networks (WANETs) with slow fading channels. We formulate the control problem as a dynamic robust game which is well-justified by two key observations: first, the shared nature of the wireless medium which inevitably cross-couples the nodes' forwarding decisions, thus binding them together as strategic players; second, the stochastic dynamics associated with the link qualities which renders the transmission costs noisy, thus motivating a robust formulation. Given the non-stationarity induced by the fading process, an online solution for the formulated game would then require an adaptive procedure capable of both convergence to and tracking strategic equilibria as the environment changes. To this end, we deploy the strategic and non-stationary learning algorithm of *regret-tracking*, the temporally-adaptive variant of the celebrated *regret-matching* algorithm, to guarantee the emergence and active tracking of the correlated equilibria (CE) in the dynamic robust forwarding game. We also make provision for exploiting the channel state information (CSI), when available, to enhance the convergence speed of the learning algorithm by conducting an accurate (transmission) cost estimation. This cost estimate can basically serve as a model which spares the algorithm from extra action exploration, thus rendering the learning process more sample-efficient. Simulation results reveal that our proposed solution excels in terms of both the number of transmissions and load distribution while also maintaining near perfect delivery ratio, especially in dense crowded environments.

**Keywords:** Slow Fading; Broadcasting; Wireless Ad-Hoc Networks; Regret Tracking; Dynamic Robust Games.

## 1. Introduction

Network-wide broadcast is a fundamental primitive in wireless ad hoc networks (WANETs) as well as an enabling mechanism for the route discovery phase of almost all on-demand routing protocols (e.g., AODV[1]). It has been shown that naive broadcast solutions such as basic flooding will give rise to the notorious broadcast storm problem [2] as a result of which the network's normal operation will be paralyzed with a huge volume of redundant messages in transit. This observation has spawned a large wave of publications on efficient forwarding whose origins almost date back to the advent of the ad hoc networks themselves [3-5]. The bulk of the literature in this area still consists of methods which work under the assumption that the links are perfectly reliable at all times. Within this mindset and assuming centralized knowledge of the topology, broadcasting can essentially be formulated as a classical constrained optimization problem with the objective of minimizing the number of transmissions while at the same time guaranteeing 100% delivery ratio [6]. Under the link reliability assumption, there also exist many graph-theoretic approaches for constructing efficient communication substrates over which broadcast messages can be thoroughly disseminated. Spanning tree and connected dominating set constructs have been at the forefront in this direction for which many approximation algorithmic techniques have been proposed to work around the issue of computational complexity [7-10]. Prior art is also ripe with a wide variety of distributed sub-optimal heuristic algorithms which draw on one (or two) hop topological knowledge for making on-the fly forwarding decisions [11-13]. Finally, when managing many-to-all broadcasts, i.e., when multiple sources tend to broadcast messages in the network, a recent trend has been to use network coding [14-17]. Instead of relaying received packets separately, network coding enables nodes to combine several packets and send out fewer combined/coded packets.

---

[*] Corresponding author: Mehdi Dehghan, Associate professor, e-mail address: dehghan@aut.ac.ir, URL: http://ceit.aut.ac.ir/~dehghan. Postal address: Mobile and Wireless Networks Lab, Department of Computer Engineering and Information Technology, Amirkabir University of Technology, PO Box 15875-4413, 424 Hafez Avenue, Tehran, Iran. Tel: +98-2164542749, Fax: +98-2166495521.

However, forwarding control in a real networking context may largely deviate from ideal abstract models given the influence of noise, fading and interference on wireless links which can give rise to unmanageable outbursts of re-transmissions. Research on supporting broadcast in the presence of unreliable links has mainly revolved around devising efficient acknowledgement (ACK) schemes [18] or alternatively, introducing redundancy into the set of forwarders [19]. There have also been a few attempts which incorporate the expected costs associated with fixed error probabilities of the outgoing links into the broadcast substrate construction [6,20]. Though applicable to the lossy link model scenarios, the existing methods are either centralized [21] or lack a principled basis to explicitly factor the stochastic dynamics associated with variable link qualities into the problem formulation. In essence, to capture the realistic effect posed by channel dynamics on message propagation gives rise to a decentralized stochastic control problem which has not been methodically investigated before.

In a departure from the prior art, this paper addresses broadcasting in multi-hop wireless networks with a realistic physical layer. We explicitly account for the variable quality of the links by assuming fading channels with slowly evolving SNR values. Hence, one can assume that the typical stochastic dynamics dealt with in this paper manifest themselves as a result of distance-related attenuation or scattering due to obstacles and terrain conditions, and evolve over moderately long time scale compared to the baseband signal variations and are associated with low Doppler spread [22].

Under slow fading, the cost incurred by a forwarding node is associated with the number of (re)transmissions it takes to deliver a given message to its neighbors. As the link reliability is changing dynamically, the forwarding costs are generally random and depend on instantaneous channel conditions. Hence, an integral facet of the broadcast control solution would be to explicitly provide for robustness against such uncertainty. On the other hand, over the course of a broadcast, the wireless medium is shared via many potential forwarders with possibly overlapping neighbor sets that incur different transmission costs for their forwarding attempts. An uncoordinated broadcast effort may not only trigger superfluous forwarding but it can also fail to proactively utilize links with high quality and avoid those in poor conditions. Therefore, forwarding control in a WANET gives rise to an inherently strategic setting, given the spatial dependency between a node and its neighbors and the resultant cross-impact of their decisions on their mutual performance.

With these understandings, in this paper, we address the channel-adaptive broadcast coordination problem in WANETs by making the following contributions:

- We come up with a game-theoretic-formulation of the forwarding control problem using the framework of *dynamic robust games* [23]. Dynamic robust games formalize repeated interactions of a set of strategic players in uncertain (noisy) environments. The robust game specification features a state variable which is of random nature and evolves over the stages of the play. Hence, by incorporating the wireless channel states into the game definition, we can explicitly cater for the noisy transmission costs incurred by the forwarding nodes. Furthermore, dynamic robust games are of incomplete information, and thus impose minimal informational assumptions on the part of the players. This would prove a desirable property as it directly translates into minimal control message overhead for distributed node coordination. In effect, we demonstrate that only one-hop ACK messages will suffice to build and maintain the nodes' forwarding strategies.

- In our forwarding game, at every broadcast interval, each node's decision is a choice between whether or not to forward the current message in transit. Given the coupling between the nodes' forwarding decisions, a real-time coordination mechanism is needed to form a global consensus (equilibrium) across the network. Furthermore, the nodes' collective forwarding behavior should be agile in tracking this consensus as the forwarding utilities change due to slow fading. To this end, we deploy the game-theoretic learning algorithm of *regret-tracking* [24,25], the temporally-adaptive variant of the celebrated *regret-matching* algorithm [26,27], to guarantee the emergence and active tracking of correlated equilibria (CE) [28] between the forwarding strategies of the network participants.

- Our solution, henceforth referred to as "regret-tracking broadcast" (RTB), is particularly suited for scenarios where no prior knowledge of the fading process and network topology is available. More specifically, in learning-theoretic parlance, RTB works in the pure bandit feedback setting [29] in which only the noisy numerical value of the utility for the actually implemented forwarding decision is perceived at each decision period. However, when channel state information (CSI) exists, an important aspect in our design is to present a model-based variant of RTB which can exploit CSI for obtaining an accurate estimate of the forwarding costs. This cost estimate essentially makes the nodes' utility functions partly known, and as shown empirically through experiments, the learning process can be expedited using this semi-bandit feedback.

- Experimental and comparative results demonstrate RTB's superior performance in terms of both the number of transmissions and load distribution while also maintaining near perfect delivery ratio in the presence of time-varying link qualities.

The rest of the paper is organized as follows: Section 2 reviews the prior art in game-theoretic and learning-based forwarding for WANETs. In section 3, we present our game-theoretic formulation of the channel-adaptive broadcast coordination problem. In 4, we first provide a brief background on the regret-based machinery for strategic learning, and will subsequently motivate our regret-tracking-based solution of the formulated game. This section continues with outlining the basic form of the RTB algorithm which is then followed by the exposition of its model-free and model-based variants in two accompanying subsections. Section 5 is dedicated to the comparative numerical evaluation of the RTB algorithm. The paper ends with a concluding epilogue.

## 2. Game-Theoretic and Learning-Based Forwarding: A Review of the Prior Art

In this section, we review related work on forwarding in WANETs. Forwarding (both unicast and broadcast) has been approached with a variety of techniques and toolboxes in the literature. We deliberately avoid giving an exhaustive overview of the many well-known and long-established solution techniques for broadcasting. Such reviews are routinely included in nearly every paper that addresses this problem (e.g., see [6]). With this in mind, we primarily devote this section to the review of game-theoretic and learning-based forwarding schemes. We believe this makes our review less stereotypical and more specialized for an interested reader.

Major prior art game-theoretic and (or) learning-based forwarding schemes belong to the realm of unicast transmissions. The main rationale for resorting to game-theoretic or socioeconomic models in devising unicast forwarding solutions is to induce cooperative behavior in scenarios where nodes (or device holders) are not operating under the control of the same authority, and may thus exhibit selfish behavior to save their limited resources. In fact, this is the direct ramification of granting autonomy to network nodes for the merits of decentralization or self-configuration. In such autonomic settings, the prevalent trend in incentivizing cooperation has been to deploy reputation- or credential-based schemes [30], or alternatively to pre-configure the forwarding task based on some static or offline computational mechanism design [31]. For instance, the study in [32] reviews a number of repeated-game-theoretic strategies (e.g., "tit-for-tat" or "grim-trigger") which organize for the network's forwarding operation to proceed on the basis of a pre-conceived Nash equilibrium (NE). The more recent trend in this direction is the study of cooperation development in ad-hoc networks by leveraging on ideas from evolutionary game theory (EGT) [33]. An interesting case is reported in [34] where the authors conduct an EGT-based analysis to determine the impact of the network size as well as the types of participating nodes on the development of cooperation. Also, the authors in [35] adopt asymmetric multi-community EGT to formulate competition among nodes in sparse VANETs. Our game-theoretic formulation, however, targets broadcast transmission scenarios. Accordingly, the application of games (or more specifically, dynamic games) in this paper is mainly of control-theoretic interest, i.e., to use games as an efficient toolbox for exerting decentralized control [36] and coordination over the network rather than as a means of cooperation stimulation.

In the context of broadcast-type forwarding, a number of studies (e.g., [37], [38], [39], and [40]) exploit some classical game settings such as Diekman's "*volunteer's dilemma*" [41] or Arthur's "*Santa Fe bar problem* (SFBP)" [42] to strike coordination between the nodes' rebroadcasting decisions. The "*volunteer's dilemma*" models a situation in public economics where each player faces the decision of either making a small sacrifice from which all will benefit, or freeriding. The "*forwarding dilemma game* (FDG)" in [37] is an adoption of Diekman's voluntary contribution problem to the game of forwarding or not forwarding a flooding packet in MANETs. The game analysis in [37] is offline, and the nodes set their forwarding probability according to a parameterized symmetric mixed NE of the game. The NE's parameters, however, are assumed to be either a priori-known or be derived from simulation experiments, which limits its practicability. Similar FDG-like systems have been introduced in [38] and [39] for VANETs and wireless sensor networks (WSNs), respectively. The SFBP, on the other hand, typifies scenarios where a congested resource, a bar in the seminal article [42], is shared by a set of agents, i.e., the bar customers. The customers enjoy their night at the bar only if it is not overcrowded. The authors in [40] propose an SFBP-based forwarding model for WSNs in which the nodes should make their forwarding decisions based on both the congestion level of their channels (i.e., the number of concurrent accesses on the channel) and their remaining energy; however, they skip derivation of the game equilibrium, and propose instead a heuristic algorithm where the nodes adjust the parameters of their own utility functions depending on their current energy level and a simple threshold-based estimation of the channel congestion.

The two more closely related studies to our work are [6, 43]. In [43], a normal-form game with known utilities has been used to model the problem in which the nodes will set their forwarding probability according to their part in the mixed NE of the game. However, given the anti-coordination nature of the game conceived in [43], the authors could have guaranteed a higher social welfare by solving the game for a (private) CE instead, while still preserving the fairness property of the mixed NE. Also, one-shot games are hardly a realistic model to capture the important aspects of the problems in dynamic settings (such as ad-hoc networks) with random and time-varying system parameters. The work in [6] presents a distributed scheme for constructing a broadcast tree over unreliable links using the notion of exact potential games [44]. The game is played by the descendants of the internal nodes and is of a cost sharing type; i.e., the cost of an edge is shared evenly by all players whose paths contain that edge, effectively directing the construction towards a spanning tree with minimum number of internal nodes who also suffer the least for their forwarding endeavor. Given the potential structure of the game in [6] and using one-hop topological knowledge, the iterative best response algorithm has been used for convergence to an NE. The broadcast tree construction in [6] depends on a precise probabilistic model of the wireless connections and the local topology of the network. In a practical setting, however, these probabilistic models have to be "learned" and "maintained." Moreover, the forwarders set in [6] is not maintained in response to variations in link qualities and hence it cannot opportunistically exploit the spatial and temporal diversity of the wireless channels across the network.

Unlike [6], in this paper, we present a structure-less broadcasting scheme which is based on a dynamic robust game played by the forwarding nodes, themselves. Dynamic robust games are just the right specification for scenarios where it is needed to capture the uncertainty associated with both the random activity of the nodes and the variability of the state of the system. We also introduce cognition to the nodes' forwarding decisions to enable proactive adaptation even when the game is of incomplete information and the environment dynamics (i.e., the channel fading process) is unknown; our solution is based on an adaptive regret-based procedure [24-27] and works within the confines of bounded rationality, a practical assumption consistent with the limited capabilities of the wireless nodes.

## 3. Broadcasting in Slow Fading WANETs: A Game-Theoretic Formulation

In this section, we formalize the broadcast coordination problem by proposing a game-theoretic formulation which readily captures the coupling between the nodes' forwarding decisions. In particular, each node's decision is simply assumed to be a choice between forward or do not forward a message at a given time. Also, the individual gain obtained by each node in the game is defined to be its local coverage ratio. Given the overlap between the transmission ranges, each node's gain is affected not only by its own decision but also by the decision of other potential forwarders. A node's forwarding cost, on the other hand, is taken to be the expected total number of transmission attempts until successful delivery to its one-hop neighbors. Under slow fading, the forwarding cost depends on the node's channel states which evolve randomly with time. Hence, the nodes' forwarding utilities should be defined in a (channel) state-dependent manner. This rules out static games as a suitable formalism in our setting, since these games are defined for a single-shot play and do not account for the evolution of players' utilities over time. To capture the uncertainty associated with the channel states, we use a dynamic robust game specification [c.f., 23, chapters 4 and 7]. Dynamic robust game is a generic term to refer to game-theoretic formulations which capture multi-stage interactions of a set of agents in uncertain (noisy) environments. Basically, the players' utilities in a robust game specification are also functions of some state variables which are of random nature and evolve over the stages of the play independently of the players' actions. This readily corresponds to our case in that the forwarding game is also modulated by current channel states whose evolution is governed solely by the slow fading process. More formally, the dynamic robust game for the forwarding control problem is a quadruple $\mathcal{G} = (\mathcal{N}, (A_i)_{i \in \mathcal{N}}, \mathcal{S}, (u_i(.))_{i \in \mathcal{N}})$, where $\mathcal{N}$ is the set of players, $A_i$ denotes the set of actions available to player $i$, $\mathcal{S}$ represents the set of states of the game, and $u_i(.)$ is player $i$'s utility function. In what follows, we give a detailed description of the components of the game $\mathcal{G}$ in the form of subsections 3.1 to 3.4. Then, in 3.5, we give a formal definition of the forwarding game's objective.

### 3.1 Set of Players in the Forwarding Game

Without loss of generality, we assume that the broadcast flow emanates from a single source node which periodically sends out critical broadcast messages to be diffused across the network. The dissemination process should be carried out in a reliable fashion so that if any forwarding attempt fails, re-transmissions are in order. Over the course of a single network-wide message dissemination, every local ensemble of nodes that are currently in hold of a fresh copy of a broadcast message $\mathcal{M}_{seq}$ with sequence number $seq$ form a set of strategic players provided that their immediate neighbor sets intersect (see Fig. 1). The fact that a node $i$'s payoff is only affected by a subset of other nodes makes our scenario an instance of a graphical game [45], a notion which is also well-suited to wireless ad hoc settings. With a slight abuse of notation, the symbol $\mathcal{N} = \{1, 2, \ldots, |\mathcal{N}|\}$ denotes a representative set of such players. It is noteworthy that a node $i$ in the forwarding game does not need to be explicitly aware of its fellow players; instead, it suffices to only infer its actually realized payoff at each stage.

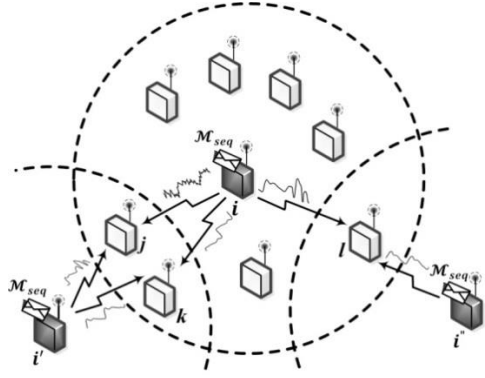

**Fig. 1.** Nodes $i, i'$ and $i''$ are currently in hold of the broadcast message $\mathcal{M}_{seq}$ and may choose to forward or not. Node $i$ has $j$ and $k$ as common neighbors with $i'$ and has $l$ in common with $i''$. The links $ij, ik, il, i'j, i'k$ and $i''l$ may differ in terms of their instantaneous signal quality.

### 3.2 Forwarding Actions and Strategies

The action space $A_i = \{0, 1\}$ for every node $i$ simply consists of two choices: to either $'drop' \triangleq 0$ the current message $\mathcal{M}_{seq}$ or $'forward' \triangleq 1$ it to a neighboring node with a possibly limited number of retries. The game proceeds with imperfect monitoring, i.e., a node does not need to observe the actions taken by the other potential forwarders. Let $\mathcal{A} = \times_{i=1}^{|\mathcal{N}|} A_i$ be the joint action space of all nodes in $\mathcal{N}$. We denote by $\boldsymbol{\pi} \in \Delta(\mathcal{A})$ a joint probability distribution over $\mathcal{A}$.

### 3.3 Forwarding Game States

We use $s_i^n(j)$ to denote the channel quality state of the link connecting the $i$-th node to its immediate neighbor $j$ at the $n$-th stage of the game. In general, the evolution of fading channels can be modeled as a finite state Markov chain (FSMC) (e.g., see [46]). In this model, the SNR range is discretized into $K$ distinct regions and then mapped into a finite-state space: $S_i(j) = \{s_1, s_2, \ldots, s_K\}, \forall i \in \mathcal{N}, \forall j \in N_i$, with $N_i$ being the immediate neighbor set of the $i$-th node. More precisely, suppose a set $\Gamma$ of $K + 1$ SNR thresholds: $\Gamma = \{\Gamma_1 = 0, \Gamma_2, \ldots, \Gamma_{K+1} = \infty\}$. Assume $s_i^n(j) = \gamma$, where $\gamma$ is the instantaneous SNR associated with the link $ij$. If $\gamma$ satisfies $\Gamma_k \leq \gamma < \Gamma_{k+1}$, the $ij$ channel is said to be in state $s_k$. In FSMC model for slow fading, we assume that $\gamma$ evolves slowly with time; i.e., at time $n + 1$, it is highly likely that $\gamma$ stays within the same region as it was at time $n$, but there is also a slight chance that it transitions to other regions. When a node probes the channel, the steady-state probability of being in the $k$-th state is given by :

$$v_k = \int_{\Gamma_k}^{\Gamma_{k+1}} g(\gamma) d\gamma, \quad k = 1, 2, \ldots, K. \quad \quad (3\text{-}1)$$

where, $g(\gamma)$ is the probability density function (PDF) of $\gamma$. Also, let $S_i = [S_i(j)]_{j \in N_i}$, then, we use $\boldsymbol{S} = \times_{i=1}^{|N|} S_i$ to denote the channel state composition over all the nodes. The learning scheme discussed in the next section accommodates for two variants of the forwarding game: the zero-knowledge scenario where CSI $\boldsymbol{s}_i^n = [s_i^n(j)]_{j \in N_i}$ is unavailable to the nodes and the case that CSI can be exploited to speed up the learning process. However, in both variants, the law of transitions between states is assumed unknown.

### 3.4 Individual Node Utilities

The instantaneous utility $u_i^n$ of a node $i$ at stage $n$ of the forwarding game is a random variable comprised of a reward and a cost component. When $i$ chooses to forward a message, the accrued reward depends on whether or not the maximum number of trials is capped in the actual protocol implementation; two reliability regimes can be envisaged: *semi-reliable* or *reliable* forwarding. The semi-reliability regime limits the number of re-transmissions to a given maximum in case extremely high error rates are being experienced like, for example, when a packet is hit by a deep fade. When CSI is available, a primitive scheme to implement semi-reliability is to approximate the number $c^*$ of trials needed to deliver the packet with a desired minimum probability $\delta \in (0,1)$. Technically, the number $c^*$ is the $\delta$-quantile of the random variable indicating the number of trials needed by a node to successfully transmit a message over its outgoing links. With $F$ being the probability distribution of the number of trials, $c^*$ is given by: $c^* = F^{-1}(\delta) = \min\{c \in \mathbb{N} : F(c) \geq \delta\}$.

The reliable regime, on the other hand, is based on the assumption that when a node decides to take part in the forwarding operation, it keeps re-transmitting until either the message $\mathcal{M}_{seq}$ is successfully delivered to all its next-hop neighbors or the next message $\mathcal{M}_{seq+1}$ is received by $i$, marking the extinction of $\mathcal{M}_{seq}$. There is, however, a price to pay for such persistence which we capture by $c_i^n$, denoting the actual number of re-transmissions caused by physical layer errors.

Also, let $|\acute{N}_i^n|$ be the cardinality of the set of $i$'s covered neighbors for $a_i^n = 1$; $r_i^n = \frac{|\acute{N}_i^n|}{|N_i|}$ denotes the reward $i$ accrues for taking part in the forwarding operation at stage $n$. In case a node chooses to drop $\mathcal{M}_{seq}$, it incurs no cost, yet it might still accrue a non-zero reward $\acute{r}_i^n = \frac{|\acute{N}_i^n|}{|N_i|}$ given that a subset $\acute{N}_i^n \subseteq N_i$ of its next-hop neighbors may receive

$\mathcal{M}_{seq}$ through other forwarders. A non-forwarding node $i$ would be able to count the members of $\hat{N}_i^n$ by simply overhearing (i.e., idle listening) the Acks its next-hop neighbors send out for receiving $\mathcal{M}_{seq}$.

Now that the reward and cost components of the utility are specified, the equation in (3-2) is considered to be the instantaneous utility a node $i$ actually perceives at each stage of the reliable forwarding game. The coefficient $\alpha$ in (3-2) is a constant scaling parameter between the two parts of the utility. Ideally, $\alpha$ should be chosen according to node density to give a reasonable trade-off between delivery ratio and the forwarding cost. As can be noted, the definition of the local performance measure at each node is in line with the global objective of minimizing the number of transmissions while guaranteeing near perfect delivery ratio.

$$u_i^n = \begin{cases} r_i^n - \alpha . c_i^n, & a_i^n = 1 \\ \acute{r}_i^n, & otherwise \end{cases} \qquad (3\text{-}2)$$

### 3.5 System-Wide Objective

To achieve global coordination of the nodes' forwarding decisions, we seek correlated equilibria (CE) [28] of the forwarding game as the system-wide solution concept. Structurally, the set of CE of a game is a convex polytope of joint action probability distributions which possess an equilibrium (quiescence) property; i.e., a CE represents competitively optimal behavior between the nodes, in which the action of each node is an optimal response to the actions of other potential forwarders. Compared to Nash equilibrium (NE), the notion of CE directly considers the ability of the nodes to correlate their actions. This correlation can lead to higher performance than if each node was required to act in isolation as is the case in NE. Moreover, the convexity of the set of CE arguably allows for better fairness between the nodes [47], which is also evidenced by our simulation experiments. However, in our dynamic robust game formulation, the channel states and the nodes' forwarding utilities evolve according to the slow fading process. Hence, the set of CE of the forwarding game should also be defined in a state-dependent manner. Let $s \in \mathcal{S}$ be a global channel state. We denote by $\boldsymbol{\pi_s} \in \Delta(\mathcal{A})$ a probability distribution over the joint action space $\mathcal{A}$ for state $s$. The state-dependent set of CE of $\mathcal{G}$, denoted by $\mathbb{C}(s)$, is defined as (3-3) below [24,25]:

$$\mathbb{C}(s) \overset{\text{def}}{=} \left\{ \boldsymbol{\pi_s} : \sum_{\boldsymbol{a}_{-i} \in A_{-i}} \boldsymbol{\pi_s}(a, \boldsymbol{a}_{-i}) . [u_i(b, \boldsymbol{a}_{-i}; s_i) - u_i(a, \boldsymbol{a}_{-i}; s_i)] \leq 0, \forall a, b \in A_i, i \in \mathcal{N} \right\}, \quad (3\text{-}3)$$

in words, if a joint forwarding action $(a, \boldsymbol{a}_{-i})$ is drawn from a CE distribution $\boldsymbol{\pi_s} \in \mathbb{C}(s)$ (presumably by a trusted third party), and each node , $i \in \mathcal{N}$ is told separately its own component $a$, then it has no incentive to choose a different forwarding action $b$, because, assuming that all other nodes $i' \in \mathcal{N}\backslash\{i\}$ also obey, the suggested action $a$ is the best

                                                                                                                 

in expectation [48]. Therefore, reaching CE can be viewed as formation of a suboptimal consensus amongst the nodes' forwarding strategies.

To compute $\mathbb{C}(s)$ in the forwarding game $\mathcal{G}$, a few remarks are in order: Basically, at each stage $n$ of the game $\mathcal{G}$, a node $i \in \mathcal{N}$ would simply choose a forwarding action $a_i^n$ and receive a numerical noisy value $u_i^n$ of its utility at that stage. Hence, a player's information at each stage consists of her past own-actions and perceived own-utilities. A private history $h_i^n$ of length $n$ for node $i$ is a collection: $h_i^n = (a_i^0, u_i^0, a_i^1, u_i^1, ..., a_i^{n-1}, u_i^{n-1}) \in H_i^n \coloneqq (A_i \times \mathbb{R})^n$. In such setting, each node $i$ selects its actions autonomously according to a strategy $\sigma_i$ which is a map $\sigma_i^{n+1}: \cup_n H_i^n \to \Delta(A_i)$. In other words, $\mathcal{G}$ is a game of incomplete information and imperfect monitoring. Given this minimal amount of information available to the players, the state-dependent CE $\mathbb{C}(s)$ of $\mathcal{G}$ cannot possibly be characterized through introspection and rationalistic (pre-play) analysis. Instead, the solution is inevitably online in the sense that the rational behavior should arise naturally via live repeated interactions during which the nodes indirectly acquire a coordination signal through their realized payoffs. In other terms, the nodes iteratively craft their strategies $\sigma_i$ and update them by using their gradually built private history of the game. In the next section, we resort to recent results from the strategic learning literature [24,25] to shape the nodes' forwarding strategies in real time so that their collective behavior tracks the system-wide solution concept $\mathbb{C}(s^n)$ as it evolves under slow fading.

## 4. Cognitive Forwarding Control through Regret Tracking

In this section, we deploy an adaptive heuristic, viz. "regret tracking broadcast (RTB)" which is built on the "*regret tracking*" procedure proposed in [24,25] in order to learn the expected payoffs simultaneously with the CE strategies of the dynamic robust game defined in subsection 3.5. Before presenting RTB's pseudo-code, we first introduce its regret-based learning engine for shaping the nodes' forwarding strategies $(\sigma_i)_{i \in \mathcal{N}}$, and will subsequently motivate our regret-tracking-based design. In 4.2, we discuss two variants of RTB: one that works without the knowledge of CSI, and the other variant that exploits the availability of CSI to enhance the learning process. Finally, in 4.3, we discuss RTB's properties in terms of convergence and computational complexity.

### 4.1 Regret-Tracking Broadcast (RTB)

Consider again the binary-valued strategy space of the nodes in our forwarding game (i.e., $A_i = \{0, 1\}$). The dynamics of a game under a *regret-matching* procedure [26,27] generally proceeds as follows (See Fig. 2 for a schematic illustration): At stage $n$, each player $i \in \mathcal{N}$ perceives its utility $u_i^n$ gained from implementing its forwarding decision $a_i^n = a$. It then computes two quantities:

- First, an estimation $\hat{U}^n(1-a)$ of the average potential utility it could have obtained had it chosen the alternative action $(1-a)$ instead of $a$ in all past plays of $a$ throughout the entire history of the game; the reason for this estimation is that the direct calculation of $u(a_i^n, \boldsymbol{a}_{-i}^n; s_i^n)$ is not possible in our case given that node $i$ only perceives its instantaneous utility at each stage in the form of a numerical value. A general technique for obtaining the estimate $\hat{U}^n$ is through the notion of *proxy regrets*, first introduced in [27]. However, as we discuss in 4.2.2., there is a possibility for obtaining better quality estimates in our case.

- Second, the average of the perceived $u_i^\eta$ utilities it has actually accrued over the stages it has chosen to play $a$; i.e., over the course of $\{\eta \le n : a_i^\eta = a\}$. We denote the value of this average at stage $n$ by $U^n(a)$ which is calculated as follows:
$$U^n(a) \coloneqq \frac{1}{n} \sum_{\eta \le n : a_i^\eta = a} u_i^\eta. \quad (4\text{-}1)$$

Let $[.]^+$ denote $\max\{.,0\}$. The difference $Q_i^n(a, 1-a) = \left[\hat{U}^n(1-a) - U^n(a)\right]^+$ is technically called the *regret* for not having played $(1-a)$ instead of $a$ over the course of the stages $\eta \le n : a_i^\eta = a$. In other words, $Q_i^n(a, 1-a)$ simply denotes the increase, if any, in the average payoff that would result if all past plays of action $a$ were to be replaced by action $(1-a)$, and everything else remained unaltered [26,27]. The player $i$ then switches to action $(1-a)$ with a probability $prob(1-a)$ proportional to $Q_i^n(a, 1-a)$ and sticks with $a$ by $1 - prob(1-a)$. The game moves on to stage $n+1$ and the process repeats.

It has been shown in [26,27] that if all players follow the update rule prescribed by the aforementioned regret-based procedure, their joint empirical frequency of play asymptotically converges to the set of correlated equilibria of the game. However, as described in Section 3.5, the equilibrium set $\mathbb{C}$ in our forwarding game is state-dependent and time-varying. Therefore, the mere notion of convergence does not suffice. It is further required that this set be tracked in time as the channel states change due to slow fading. We now elaborate on the key change that has to be made in updating a node's regret-based forwarding strategy to account for the slow fading effect. Our discussion here is based on the theoretical results in [24,25]. We only go through the intuition behind the main idea in tracking $\mathbb{C}(\boldsymbol{s}^n)$ and refer the reader to [24,25] for technical exposition. In the standard regret-based scheme we just described, the decisions of each player are based on the *uniform* average history of all past observed utilities (note the factor $\frac{1}{n}$ in equation (4-1)). Such uniform treatment of the obtained utilities is not desirable in our setting since the fading process causes the channel states and thus the transmission success probabilities evolve over time. As a result, the utilities gained by a forwarding node is essentially noisy and their expected values may vary every once in a while. Hence, the nodes should keep a perpetual state of readiness for temporal variations in their expected utilities. This observation directs us to use a temporally adaptive variant of the *regret matching* procedure, the so-called *regret tracking* algorithm [24,25]. In regret-tracking, the average utility is computed in a discounted manner to value more recent utilities higher than more distant utilities. In particular, a constant discount factor $\varepsilon$ is used to introduce exponential forgetting of the past and to permit tracking of a slowly time-varying environment. With this change, the calculation of the average discounted actual utility gained from implementing a given action $a$ throughout the stages $\{\eta \le n: a_i^\eta = a\}$ follows equation (4-2):

$$U^n(a) := \sum_{\eta \le n: a_i^\eta = a} \varepsilon(1-\varepsilon)^{n-\eta} u_i^\eta, \quad 0 < \varepsilon \ll 1 \qquad (4\text{-}2)$$

Using regret-tracking updates, the nodes learn and maintain their part in a correlated equilibrium of the forwarding game; however, unlike the almost-sure convergence of the classical regret matching, here, convergence to the set of correlated equilibria takes place in a weaker sense. We reiterate this result more formally in Section 4.3 and refer the reader to [24-27] for extensive discussion.

Now that a complete picture of each node's learning engine is described, we may present the complete pseudo-code of our regret-tracking-based broadcast management algorithm (RTB). Of particular note is that since in RTB a node does not need to explicitly monitor the others' actions, no particular synchronization mechanism is required between the participants. This relieves the algorithm from the exchange of signaling messages given that it only suffices to have an observation of the individual utilities per learning iteration. We are thus able to present RTB in an asynchronous event-driven style (See Algorithm 1 for pseudo-code and Table 1 for symbols and definitions).

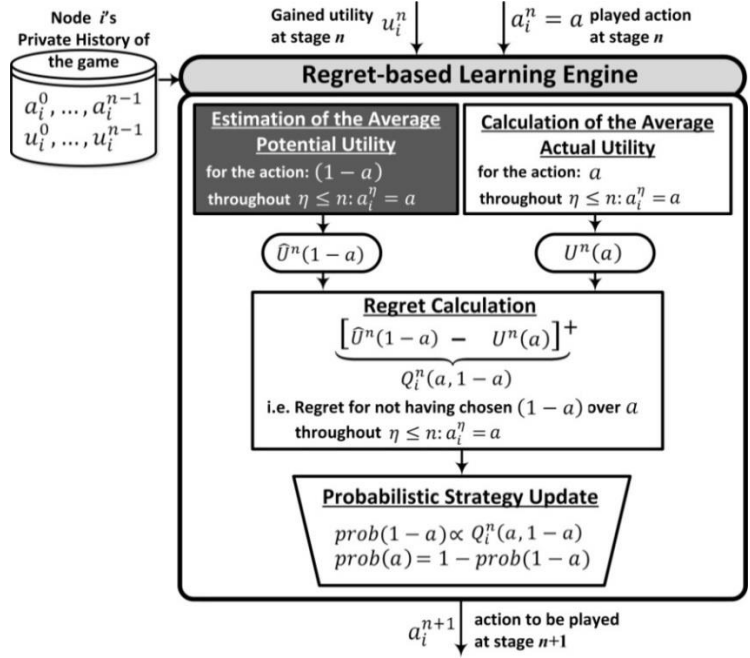

**Fig. 2.** Schematic of the regret-based machinery for strategic learning in a binary-valued strategy space. The dark-filled block (i.e., estimation of the average potential utility) constitutes the subject of Section 4.2.

**Table 1.** Notations Used in Regret Tracking Broadcast (RTB) Algorithm.

| Symbol | Definition |
|---|---|
| $\mathcal{M}_{seq}$ | fresh copy of the $seq$-th broadcast message delivered to node $i$ |
| $N_i$ | node $i$'s neighbor set |
| $\acute{N}_i$ | node $i$'s covered set ($\acute{N}(i) \subseteq N(i)$) |
| $\boldsymbol{\sigma}_i^n$ | node $i$'s forwarding strategy at stage $n$ |
| $a_i^n$ | node $i$'s selected action at stage $n$ |
| $c$ | number of (re)transmissions made by node $i$ at stage $n$ |
| $\alpha$ | a constant scaling factor |
| $u_i^n$ | instantaneous utility node $i$ actually receives at stage $n$ |
| $U^n(a)$ | actual (weighted) average utility for playing action $a$ |
| $\bar{U}^n(1-a)$ | estimated (weighted) average utility for playing $a$'s alternate action |
| $\varepsilon$ | constant discount factor, $0 < \varepsilon \ll 1$ |
| $\delta$ | exploration factor |
| $\mu$ | normalization constant (update inertia), $\mu > Q_i^n(a, 1-a)$ |
| $Q_i^n(a,b)$ | node $i$'s *regret* for not having played $b$ instead of $a$ |

- In the beginning, each node $i \in \mathcal{N}$ has a random initial action $a_i^0 \in \{0,1\}$, a zero *regret* value $Q_i^0 = 0$, empty covered set $\acute{N}(i) = \emptyset$, and zero (re)transmission count $c = 0$.

- Upon reception of a fresh copy of a broadcast message $\mathcal{M}_{seq}$ with sequence number $seq$, each node $i \in \mathcal{N}$ first fires an event indicating the expiration of its previously handled message $\mathcal{M}_{seq-1}$ (line 1). Processing the $\mathcal{M}_{seq-1}\_EXPIRED$ event provides the opportunity to update the parameters of the learning engine. The numerical value of $i$'s payoff $u_i^n$ for $\mathcal{M}_{seq-1}$ is computed in line 13 using equation (3-2). Line 14 calculates the estimate $\bar{U}^n(1-a)$, the specifics of which is the subject of discussion in section 4.2. Lines 15 to 17 are essentially the standard update routine for regret-based learning (see Fig. 2). In line 17, taking the minimum with $\frac{1}{|A_i|}$ guarantees that the assigned probabilities form a valid probability measure over $A_i = \{0, 1\}$. The second term forms a uniform distribution on $A_i$ (with probability $\delta$) and can be interpreted as the "exploration". Exploration is necessary in cases such as ours where nodes continuously learn their utility functions and ensures both actions being played with a non-zero chance [27]. Lines 4 to 8 correspond to the case when the node has chosen to forward $\mathcal{M}_{seq}$. It basically keeps sending $\mathcal{M}_{seq}$ until either all $j \in N_i$ are covered (by $i$ or others) or the next message is received by $i$. The number of transmissions made by $i$ is tracked by the counter $c$, which is a random variable whose realization depends on the channel states at time $n$. The firing of the event $ACK\_MESSAGE\_RECEIVED\_OR\_OVERHEARD$ notifies $i$ that $\mathcal{M}_{seq}$ has been received by some $j \in N_i$ either through $i$ or other forwarders. It allows $i$ to compute its local delivery ratio $\frac{|\acute{N}_i|}{|N_i|}$ which is also a random variable whose realization depends on $i$'s decision as well as the decisions of its fellow players'. In case a node chooses to drop $\mathcal{M}_{seq}$ (line 10), it just idly listens to the medium to overhear the Acks from $j \in N_i$.

---

**Algorithm 1.** Regret Tracking Broadcast (RTB) Algorithm.

---

**Initialization**:

$\acute{N}_i := \emptyset$; $c := 0$; $\sigma_i^0(0) := \sigma_i^0(1) := \frac{1}{|A_i|}$; $n := 0$;

**begin**

    **case** (*event*) **do**

        $BROADCAST\_MESSAGE\_RECEIVED$: // Received a fresh copy of $\mathcal{M}_{seq}$ from a neighboring node.

1:      Fire $\mathcal{M}_{seq-1}\_EXPIRED$;   // Node $i$ fires an event indicating the expiration of previous message $\mathcal{M}_{seq-1}$.

2:      $handled(\mathcal{M}_{seq}) = \textbf{true}$;   // Node $i$ sets a private flag indicating it is going to handle $\mathcal{M}_{seq}$.

3:      Choose $a_i^n = a$ with probability $\sigma_i^n(a)$;

4:      **if** $(a == 1)$ **then**

5:          **while** $(\acute{N}_i \neq N_i)$ **do**

6:              Broadcast $\mathcal{M}_{seq}$;

7:              $c = c + 1$;

8:          **end while**

9:      **else**

10:         /* idle listening… */

11:      **end if**

        $\mathcal{M}_{seq-1}\_EXPIRED$: // $\mathcal{M}_{seq-1}$ expires when $\mathcal{M}_{seq}$ is received.

12:      **if** $(handled(\mathcal{M}_{seq-1}))$ **then**   // Node $i$ updates its parameters only if it has handled $\mathcal{M}_{seq-1}$.

13:         $u_i^n := \frac{|\acute{N}_i|}{|N_i|} - \alpha.c$;

14:         Calculate the estimate $\bar{U}^n(1-a)$;   // The specifics of this estimation is discussed in subsection 4.2.

15:         $U^n(a) := \sum_{\eta \leq n: a_i^\eta = a} \varepsilon(1-\varepsilon)^{n-\eta} u_i^\eta$;

16:         $Q_i^n(a, 1-a) := \left[\bar{U}^n(1-a) - U^n(a)\right]^+$;

17:         $\sigma_i^{n+1}(1-a) := (1-\delta)\min\left\{\frac{Q_i^n(a, 1-a)}{\mu}, \frac{1}{|A_i|}\right\} + \frac{\delta}{|A_i|}$; $\sigma_i^{n+1}(a) := 1 - \sigma_i^{n+1}(1-a)$;   // Update forwarding probability.

18:         $\acute{N}_i := \emptyset$; $c := 0$; // Reset covered set and (re)transmission number for the next stage.

19:         $n := n + 1$. // Update the time index.

20:      **end if**

        $ACK\_MESSAGE\_RECEIVED\_OR\_OVERHEARD$: // Node $i$ is Acked or overhears an ACK for $\mathcal{M}_{seq}$ from some $j \in N_i$.

21:      $\acute{N}_i := \acute{N}_i \cup \{j\}$;

---

**end**

Thus far, we have not specified how each node obtains an estimate of its average potential utility. In the following two subsection, we give two variants of RTB which differ in their estimation of the quantity $\bar{U}^n(1-a)$.

**4.2 Estimation of the Average Potential Utility**

The RTB algorithm requires that both the forwarding and dropping decisions be somehow evaluated at each stage of the game so as to be able to update the regret values associated with a nodes' sequence of decisions. Depending on the forwarding decision made at stage $n$, a node will need to get hold of different types of information for determining the instantaneous utility its alternate choice would have yielded at the same stage:

- In case a node $i$ chooses to drop the message $\mathcal{M}_{seq}$, it needs to calculate the forwarding cost $c_i^n$ it would have actually incurred, had $i$ broadcast $\mathcal{M}_{seq}$ over its communication channels. Assuming the availability of CSI, $\boldsymbol{s}_i^n = [s_i^n(j)]_{j \in N_i}$, at the forwarding node $i$, a theoretically valid estimate of the forwarding cost would be the expected number of re-transmissions $\bar{C}_i^n(\boldsymbol{s}_i^n)$, which can be calculated using closed-form analytical expressions for BER in a given fading channel type. The enhanced learning with CSI-based cost estimation is the subject of subsection 4.2.2. However, while we can safely assume the perfect CSI at the receiver end, the inaccuracy of the channel estimation process, erroneous or obsolete feedback, and time delays or frequency offsets between the reciprocal channels may impede the sender from obtaining the perfect CSI. The estimation of the average utility $\bar{U}^n(1)$ in subsection 4.2.1 totally disregards the availability of the CSI and thus is particularly suited for scenarios where there is neither a feedback channel from the receiver to the sender nor is there a mechanism for exploiting the channel reciprocity such as in time-division duplexing systems.

- Unlike the cost component, the forwarding reward $r_i^n$ has no straightforward estimate when the message is dropped; also, in case of forwarding $\mathcal{M}_{seq}$, a node has no means to correctly determine $\acute{N}(i)$ (its covered set) given that its own transmission may result in others immaturely backing off from the forwarding endeavor. Hence, the reward components need to be estimated through the technique of *proxy regrets* discussed in 4.2.1.

**4.2.1 Zero-knowledge Learning with Proxy Regrets**

The unavailability of the information necessary for evaluating the alternate actions calls for a zero-knowledge learning scheme with bandit (or opaque) feedbacks. More specifically, a node may define a *proxy regret measure* [27] by using the utilities it has perceived thus far when it actually played the alternate action $(1-a)$ over the previous stages of the forwarding game. The calculation of the (*proxy*) regret measure $Q_i^n(a, 1-a)$ would then require that the average $\bar{U}^n(1-a)$ be estimated as follows:

$$\bar{U}^n(1-a) = \sum_{\eta \le n : a_i^\eta = (1-a)} \varepsilon(1-\varepsilon)^{n-\eta} \frac{\sigma_i^\eta(a)}{\sigma_i^\eta(1-a)} u_i^\eta. \quad (4\text{-}3)$$

In the above equation, $\boldsymbol{\sigma}_i^\eta$ denotes the play probabilities at stage $\eta$; in effect, the *proxy* regret for not having played $(1-a)$ instead of $a$ measures the difference of the average utility over the stages when $(1-a)$ was actually used and the stages when $a$ was used. The term $\frac{\sigma_i^\eta(a)}{\sigma_i^\eta(1-a)}$ normalizes the per-stage utilities $u_i^\eta$ so that the length of the respective stages would become comparable.

### 4.2.2 Enhanced Learning with CSI-based Cost Estimation

The zero-knowledge case, discussed in the previous subsection, is oblivious to the availability of CSI, and instead, relies on a rough estimation technique to approximate the average utility $\bar{U}^n(1)$ associated with a node's alternate decision to $'forward'$. However, when CSI does exist, there is room for some enhancement. In effect, to obviate the need for extra action exploration, we can exploit CSI to derive a higher quality estimate of the costs. In order to do so, we note that, given its current CSI, $\boldsymbol{s}_i^n = [s_i^n(j)]_{j \in N_i}$, and assuming BPSK modulation, a node $i$ can calculate the instantaneous BER on its links with neighboring nodes $j \in N_i$ as follows [49]:

$$BER_{ij}^n(s_i^n(j) = s_k) := \frac{\int_{\Gamma_k}^{\Gamma_{k+1}} 0.2 \times e^{-1.6\gamma} \times g(\gamma) d\gamma}{\int_{\Gamma_k}^{\Gamma_{k+1}} g(\gamma) d\gamma}, \quad (4\text{-}4)$$

Now, assuming that the broadcast message $\mathcal{M}_{seq}$ is $L$ bits long, $i$'s transmission of $\mathcal{M}_{seq}$ to $j$ would succeed with probability $p_{ij}^n$, calculated as:

$$p_{ij}^n = \left(1 - BER_{ij}^n(s_i^n(j))\right)^L. \quad (4\text{-}5)$$

To guarantee delivery, node $i$ will need to make $C_i^n(\boldsymbol{s}_i^n)$ number of (re)transmissions until $\mathcal{M}_{seq}$ successfully reaches all $j \in N_i$. $C_i^n(\boldsymbol{s}_i^n)$ is a random variable whose realization depends on $i$'s channel state in period $n$. Using link reception probabilities $p_{ij}^n$, and the derivation in [6], we may express the expected value of $C_i^n(\boldsymbol{s}_i^n)$ as follows:

$$\bar{C}_i^n(\boldsymbol{s}_i^n) = 1 + \sum_{j \in N_i} \frac{1 - p_{ij}^n}{p_{ij}^n}. \quad (4\text{-}6)$$

With $\bar{C}_i^n(\boldsymbol{s}_i^n)$ at hand, we may rewrite (4-3), i.e., the estimated average $\bar{U}^n(1 - a_i^n)$, for $a_i^n = 0$ as:

$$\bar{U}^n(1, \boldsymbol{s}_i^n) = \left[ \sum_{\eta \le n : a_i^\eta = 1} \varepsilon(1-\varepsilon)^{n-\eta} \frac{\sigma_i^\eta(0)}{\sigma_i^\eta(1)} r_i^\eta - \sum_{\eta \le n : a_i^\eta = 0} \varepsilon(1-\varepsilon)^{n-\eta} \bar{C}_i^n(\boldsymbol{s}_i^n) \right]. \quad (4\text{-}7)$$

 

In effect, we decompose the estimated average utility into its reward and penalty components, and replace the index of the cost summation to match with the stages corresponding to $'drop'$ decisions, given that now a comparable stage-by-stage estimate is available for the cost component. For ease of reference, we refer to this modified version of RTB as Enhanced-RTB.

### 4.3 Convergence and Computational Complexity

In this section, we discuss RTB's convergence and computational complexity. Similarly to [24,25], we denote by $\boldsymbol{z}_\varepsilon^n \in \Delta(\mathcal{A})$ the (empirical) average collective forwarding behavior under RTB, which can also be viewed as a diagnostic that monitors the forwarding performance of the entire network. When all nodes choose their forwarding actions $\boldsymbol{a}^n = (a_i^n)_{i\in\mathcal{N}}$ using Algorithm 1, $\boldsymbol{z}_\varepsilon^n$ can be viewed as an average or moving average frequency of play, and can be represented by the following recursion for $\forall \boldsymbol{a} \in \mathcal{A}$:

$$\boldsymbol{z}_\varepsilon^{n+1}(\boldsymbol{a}) = (1-\varepsilon).\boldsymbol{z}_\varepsilon^n(\boldsymbol{a}) + \varepsilon.\mathbb{I}_{\{\boldsymbol{a}^{n+1}=\boldsymbol{a}\}}.$$

In other terms, for a joint action $\boldsymbol{a} \in \mathcal{A}$, $z_\varepsilon^n(\boldsymbol{a}) = \sum_{\eta \leq n} \varepsilon(1-\varepsilon)^{n-\eta}.\mathbb{I}_{\{\boldsymbol{a}^\eta=\boldsymbol{a}\}}$, where $\mathbb{I}_{\{.\}}$ is the indicator function. It has been shown in [24,25] that $\boldsymbol{z}_\varepsilon^n$ asymptotically tracks the time-evolving set of CE $\mathbb{C}(\boldsymbol{s})$ (see Eq. (3-3)). In a Markovian environment, the technical condition that guarantees this tracking result is that the underlying Markov process transitions at infrequent intervals (e.g., if the mean time between state changes is $O(1/\varepsilon)$ [24,25]). This condition is satisfied in our case, since we assumed fading evolves slower than the packet–level timescale (we elaborate more concretely about this in the simulation setup). However, in the face of higher fading rapidity, since the regret-tracking procedure underlying RTB is an instance of an adaptive filtering algorithm [50,51], if the underlying random process changes too fast, then it is not possible to keep track of the time-varying conditions. This is because the dynamics of the underlying Markov process is not explicitly accounted for in the algorithm.

As for RTB's computational complexity, note that similarly to [24,25], it is possible to compute the regret measure $Q_i^n(.,.)$ in a more efficient recursive form. This avoids having to compute $Q_i^n$ from scratch in every period. For instance, the regret update equation for RTB in the zero-knowledge case can be expressed as:

$$Q_i^n(a,b) = Q_i^{n-1}(a,b) + \varepsilon\left(\left[\frac{\sigma_i^n(a)}{\sigma_i^n(b)}u_i^n(a_i^n).\mathbb{I}_{\{a_i^n=b\}} - u_i^n(a_i^n).\mathbb{I}_{\{a_i^n=a\}}\right]^+ - Q_i^{n-1}(a,b)\right). \qquad (4-8)$$

With this modification in RTB's pseudo-code, basically, at each iteration, each node needs to perform just a few standard arithmetic operations and comparisons, along with one random number generation to take the next forwarding action $a_i^{n+1}$. It is noteworthy that in this recursive form, the parameter $\varepsilon$ can be viewed as a constant step size governing the adaptation rate of the algorithm [24,25].

## 5. Performance Evaluation

In this section, we simulate the performance of the proposed cognitive forwarding scheme for managing the dissemination of broadcast messages across slow fading channels in a wireless ad-hoc environment. We assume constant packet sizes of length equal to $L = 512$ bytes. Each forwarding node transmits at a constant power of 0.1 Watts. Although RTB does not depend on any distribution for the channel SNR $\gamma$, for the purpose of modeling, we simulate slow Rayleigh channels for each link. For a Rayleigh mode, channel SNR $\gamma$ is an exponentially distributed random variable with probability density function given by $g(\gamma) = \frac{1}{\bar{\Gamma}} e^{\frac{-\gamma}{\bar{\Gamma}}}$, where $\bar{\Gamma} = \mathbb{E}[\gamma]$ is the average SNR. We discretize the channel into eight equal probability bins, with the boundaries specified by $\{(-\infty, -8.47\text{ dB}),$ $[-8.47\text{ dB}, -5.41\text{ dB}),\ [-5.41\text{ dB}, -3.28\text{ dB}), [-3.28\text{ dB}, -1.59\text{ dB}), [-1.59\text{ dB}, -0.08\text{ dB}),\ [-0.08\text{ dB},$ $1.42\text{ dB}),\ [1.42\text{ dB}, 3.18\text{ dB}), [3.18\text{ dB}, \infty)\}$. The fixed quantized average SNR value $\bar{\gamma}_k$ for each state $s_k, k = 1, 2, \dots, K$ then becomes $\bar{\gamma}_k = (v_k)^{-1} \int_{\Gamma_k}^{\Gamma_{k+1}} \gamma g(\gamma) d\gamma$, where following (3-1), $v_k = e^{\frac{-\Gamma_k}{\bar{\Gamma}}} - e^{\frac{-\Gamma_{k+1}}{\bar{\Gamma}}}$. Similarly to [52,53], the transition probability matrix $\left(\mathbb{P}_{k,\tilde{k}}\right)_{k,\tilde{k}=1,\dots 8}$ of the FSMC is assumed to have the following structure:

$$\mathbb{P} = \begin{bmatrix} \rho & \sigma & 0 & 0 & 0 & \cdots & 0 & \sigma \\ \sigma & \rho & \sigma & 0 & 0 & \cdots & 0 & 0 \\ 0 & \sigma & \rho & \sigma & 0 & \cdots & 0 & 0 \\ \vdots & \vdots & \vdots & \vdots & \vdots & \ddots & \vdots & \vdots \\ \sigma & 0 & 0 & 0 & 0 & \cdots & \sigma & \rho \end{bmatrix},$$

where $\rho = 1 - 2\sigma$ and $\sigma = \mathcal{O}(f_d \tau)$, with $f_d$ and $\tau$ denoting the Doppler frequency shift and packet duration time, respectively. The product $f_d \tau$ characterizes the fading speed of the channel relative to the packet length. A small $f_d \tau$ means that the channel fading rate is small. Throughout simulations, different instances of the matrix $\mathbb{P}$ are used to generate the channel data profile with the only restriction that $f_d \tau$ be in the same order of magnitude as RTB's step size parameter $\varepsilon$. The RTB algorithm works in the pure bandit setting and is thus not aware of the instantaneous CSI; however, for the sake of Enhanced-RTB, we assume that only finite CSI is fed back, and the nodes only know $\gamma$ belongs to an interval $[\Gamma_k, \Gamma_{k+1})$ instead of having the exact value. The nodes are assumed to have modulation and coding schemes that support a transmission rate of 1Mbps for all the links in the network. We assume that the nodes operate in a collision-free environment and that they periodically exchange beacon messages to maintain their one-hop neighbor sets. In the simulation runs, 50 nodes are distributed uniformly over a square region with the node density varying between 20 to 170 nodes/km$^2$. The node density is controlled by adjusting the simulation area while keeping the number of nodes fixed. Table 2 lists the simulation parameters used in our experiments.

**Table 2.** Simulation Parameters.

| Parameter | Value |
|---|---|
| number of nodes | 50 |
| node density | 20-170 nodes/km$^2$ |
| packet length | 512 bytes |
| fading | slow Rayleigh |
| transmission power | 0.1 Watts |
| transmission rate $T$ | 1 Mbps |
| modulation | BPSK |
| RTB's step size $\varepsilon$ | 0.1 |
| RTB's scaling parameter $\alpha$ | 0.1 for low density regime 0.3 for high density regime |
| packet origination rate | 10 pkts/sec |

Performance evaluation is done in terms of *delivery ratio*, *number of transmissions* and the *balance in load distribution*. For the sake of comparison, we simulate three other broadcasting schemes: *simple flooding with retransmissions* (e.g., RBAV in [54] or ACK-flooding in [55]), *multi-point relaying* (MPR) [56,57] with retransmissions, and the *game-based broadcast tree construction* (GB-BTC) scheme recently proposed in [6].

- **MPR:** MPR is a broadcasting scheme based on two-hop topological information. It is effectively implemented in the OLSR routing protocol, which is a proactive routing protocol ratified as a request for comments (RFC) in the Internet Engineering Task Force (IETF) MANET chapter [57]. Each node in the network selects a subset of its one-hop neighbor nodes, called multipoint relays (MPRs), as the forwarding node set to retransmit broadcast packets. Other nodes that are not in the MPR set can read but not re-transmit packets. The MPR set guarantees that all two-hop neighbor nodes of each node receive a copy of the packets and, therefore, all nodes in a network with reliable links can be covered without re-transmissions by every single node. In order to apply MPR to our unreliable setting, we incorporate an explicit ACK mechanism into the protocol operation so that a node retransmits a packet when it does not receive an ACK from any intended receivers in a predefined time interval. Variants of MPR with retransmission have been considered for instance in [58].

- **GB-BTC:** The GB-BTC scheme [6], also discussed briefly in section 2, uses the notion of potential games to construct, in a distributed fashion, a spanning broadcast tree with (approximately) minimum expected number of transmissions for all internal nodes. Unlike RTB, the game in [6] is a parent selection game to be played by the successors of each internal node. A node's utility for joining a parent node on a link is the negative sum of two cost components: the first component is inversely proportional to the number of nodes selecting that same parent, and the second cost component is proportional to the cost of the communication link connecting the node to that parent. However, the links' costs in GB-BTC (i.e., the expected number of (re)transmissions) are derived assuming fixed BERs, i.e., oblivious to the realistic dynamics affecting the wireless environment such as channel fading. Also, the construction procedure in [6] is a one-time task and there is no discussion on how to gracefully maintain the tree structure in response to changes. In effect, the best-response algorithm used in [6] would not converge in non-static environments [59]. Therefore, we have simulated the dissemination of broadcast messages in GB-BTC by constructing its tree using link costs corresponding to the initial CSI only.

We first investigate the dynamic behavior of RTB and Enhanced-RTB in terms of the progression of the average delivery ratio as well as the average number of transmissions over time as the nodes learn their forwarding strategies. We do the experiments for two scenarios with respect to node density: the 20 nodes/km$^2$ case as a representative for low density regime, and the case with 170 nodes/km$^2$ showcasing a high density scenario. We allow for unlimited number of (re)transmissions with the lag between subsequent broadcast messages large enough so that 100% delivery ratio is achievable by perfect delivery schemes such as by flooding in a collision-free setting. This would also be the case with MPR and GB-BTC; i.e., they also ensure perfect delivery ratio given their perfect coverage guarantee. Therefore, there is no need to run delivery ratio-wise experiments on these three methods. As for RTB and Enhanced-RTB, we plot the average overall delivery ratio over the course of 5000 network-wide broadcasts. Over this time, the channel conditions vary according to a fixed instance of the matrix $\mathbb{P}$. As shown in Fig. 3, in both algorithms, the average delivery ratio asymptotically approaches to 1. However, Enhanced-RTB is noticeably quicker in achieving high delivery ratio, thanks to its more accurate estimation of the forwarding costs. Also, as can be evidenced from Fig. 3, it is the case that in both algorithms, higher delivery ratio is achievable more rapidly when node density is higher. Fig. 4 plots the average total number of transmissions made by the forwarding nodes in the same simulation setup as in Fig. 3. Once again, it is the case that although, in the long run, Enhanced-RTB incurs the same number of transmissions as that of RTB, it stabilizes more quickly. We have also shown GB-BTC's average total number of transmissions over time. The scale of our setup is too coarse to expect to witness in action the ability (or lack thereof) to track link qualities. This is mainly because in a large neighborhood, the effects of variations in individual links tend to mostly offset each other. Couple this with the fact that in RTB's early operation, the delivery ratio is less than perfect, and therefore the average transmission count is tentatively lower. However, it is evident from the figure that in the limit, GB-BTC imposes a larger number of transmissions. This is because the internal tree nodes are oblivious to the fact that the quality of their links are likely to degrade in time, while in the meantime, there may be better candidates which stand idle instead of replacing those undergoing poor conditions. Hence, no matter how the uncertainties in link qualities play out, failure to account for these dynamics can result in lower performance.

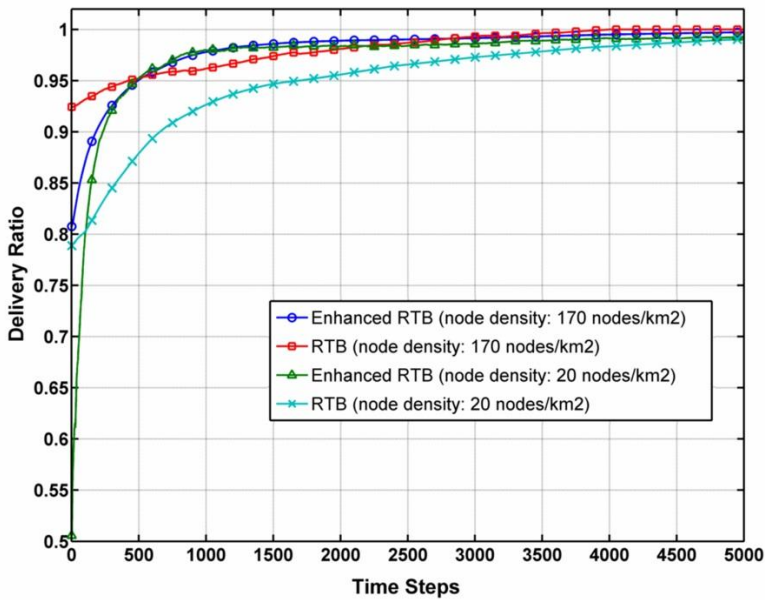

**Fig. 3.** The delivery ratio vs. time in RTB and Enhanced-RTB under dynamic channel conditions.

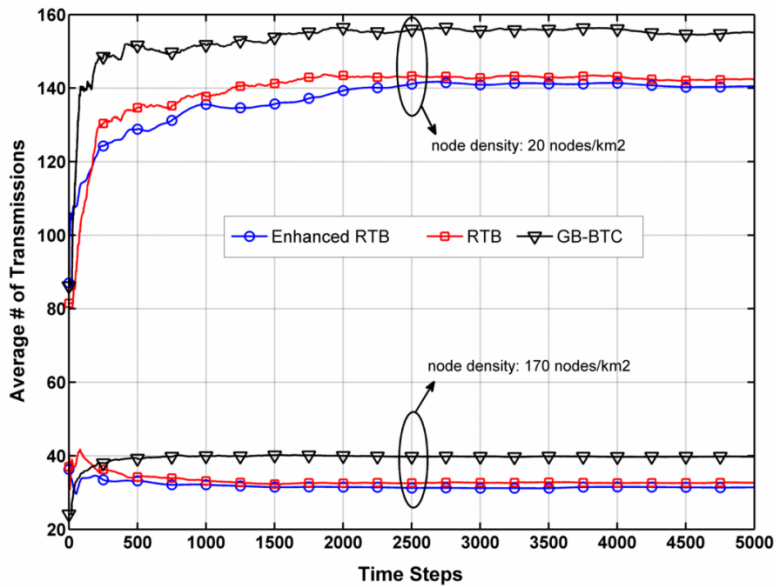

**Fig. 4.** The average total number of transmissions vs. time in RTB, Enhanced-RTB, and GB-BTC under dynamic channel conditions.

Next, more generally, we study the impact of node density on the average number of transmissions made by each of the four schemes. Each point in Fig. 5 is the average of 250 simulation runs with random topology instances. We include error bars which indicate 95% confidence that the actual average is within the range of depicted interval. The transmission count in all schemes tends to decrease as node density increases. In all cases in Fig. 5, RTB achieves perfect delivery ratio, and yet its transmission count is lower compared to other three schemes. Fig. 6 shows the average number of transmissions made by each individual node in RTB and GB-BTC. As can be seen, the distribution of load in RTB is significantly more balanced compared to that of GB-BTC. In fact, while GB-BTC's forwarding structure remains unchanged, each forwarding node in RTB consistently re-examines the value of its contribution to the overall forwarding endeavor, and once the quality of its links degrades, refrains from forwarding and instead relies more on those who enjoy higher quality links. A byproduct of these reconfigurations is the more even distribution of the broadcast load across the network. As observed from Fig. 6(b), RTB's load balancing property is much more apparent when node density is as high as 170 nodes/km², whereas GB-BTC puts the burden of forwarding on a fewer number of internal nodes.

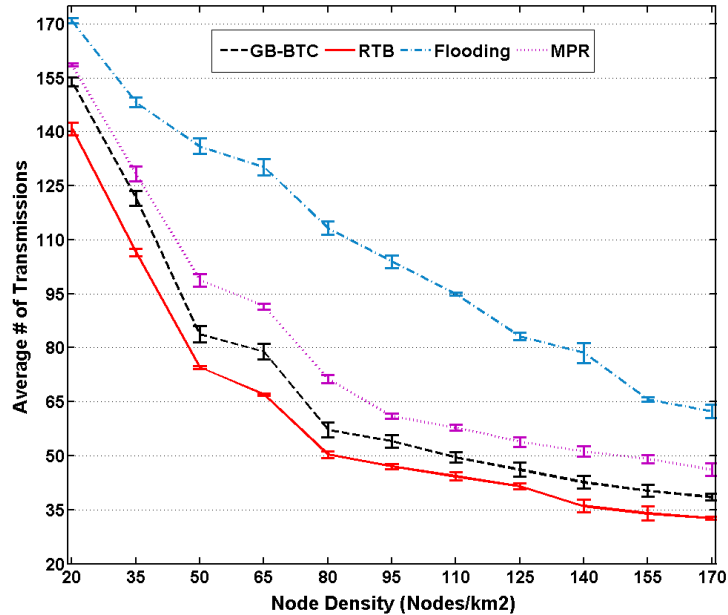

**Fig. 5.** The number of transmissions vs. various node densities under dynamic channel conditions.

We extend the simulation to consider one more scenario. We examine the performance of RTB when channel states remain static throughout the simulation; i.e., the evolution of SNR on each link is a stationary process with a constant expected value corresponding to one of the 8 regions $[\Gamma_k, \Gamma_{k+1})$. Given the stationary nature of the link variations in this case, we run RTB with a decaying step size $1/n$ to achieve an almost sure convergence to the CE set of the forwarding game. Fig. 7 shows the average total number of transmissions made by RTB and GBG-BTC in this case. RTB starts with no prior knowledge of the statistics of the link qualities, while GB-BTC has already constructed its broadcast tree with the exact knowledge of the expected SNR on each link. Although the performance margin between RTB and GB-BTC is small, but RTB is much slower to stabilize, especially when node density is low. This is also the case with delivery ratio (see Fig. 8); i.e., while GB-BTC guarantees perfect delivery from the outset, it takes a while for RTB before starting to catch up. Finally, as discussed in [6], with fixed expected SNRs, the lower bound for the average total transmission count can be obtained using a mixed integer linear program (MILP) to construct a minimum spanning tree with perfect delivery ratio. Fig. 9 illustrates the sub-optimality gap for RTB and GB-BTC under varying node densities.

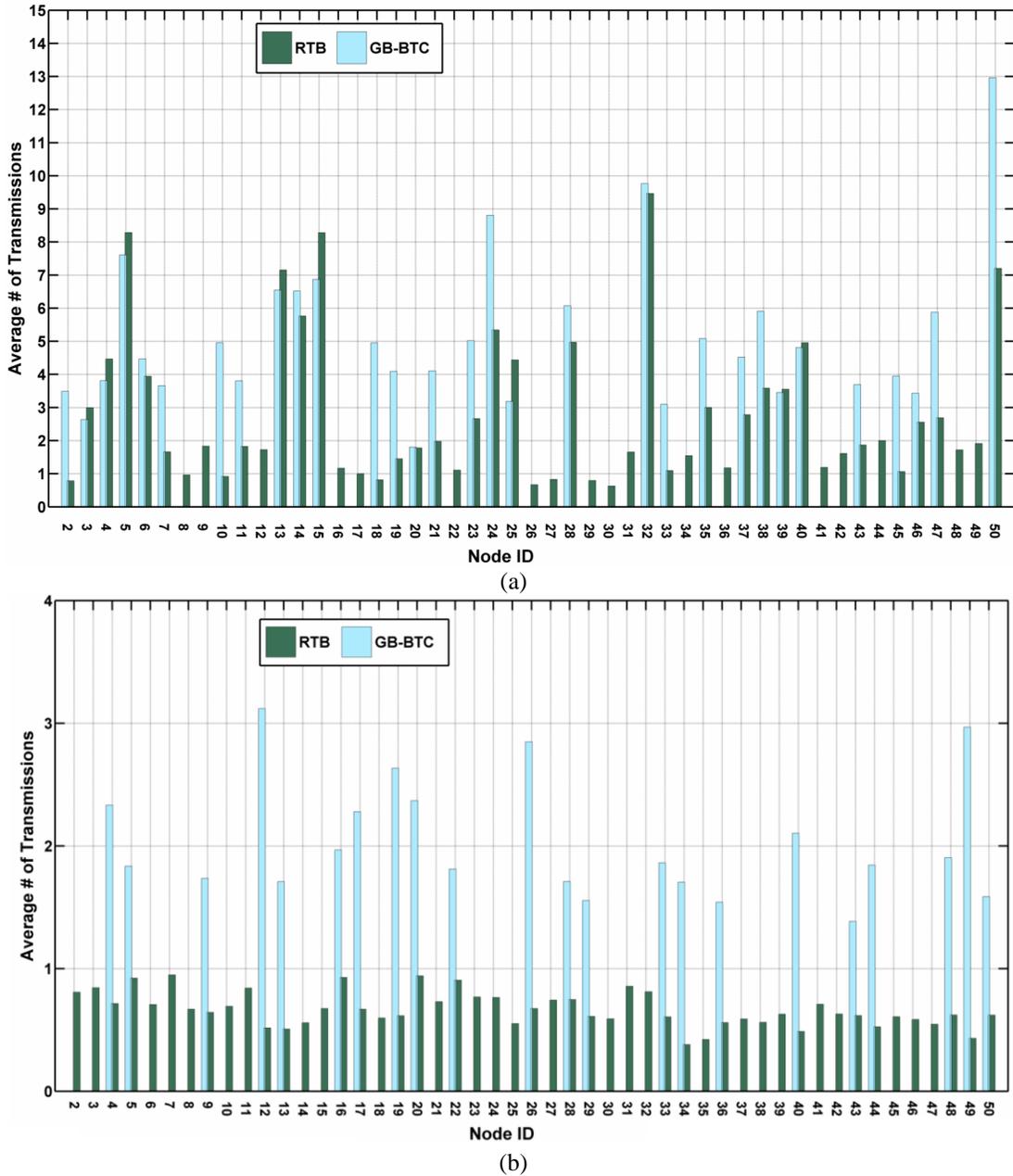

(a)

(b)

**Fig. 6.** Broadcast flow distribution in RTB and GB-BTC under dynamic channel conditions; (a): node density is 20 nodes/km$^2$. (b): node density is 170 nodes/km$^2$.

                                      

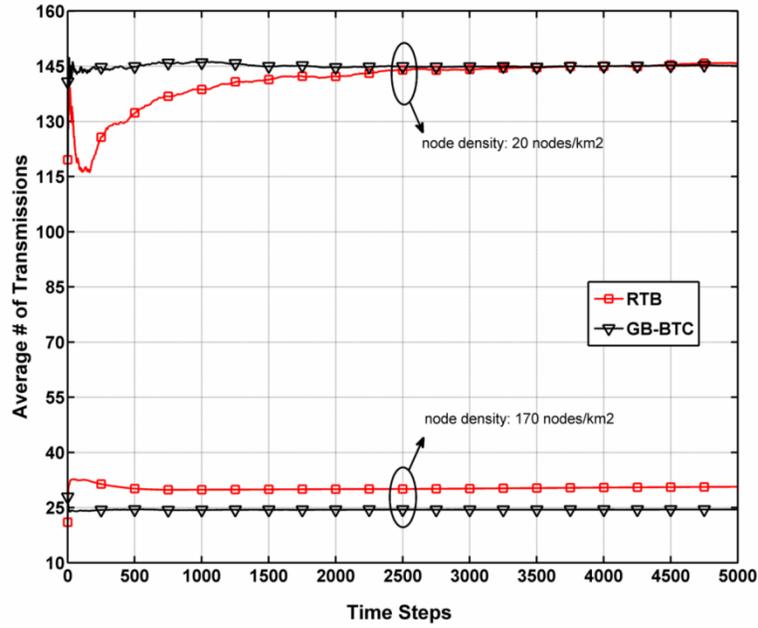

**Fig. 7.** The average total number of transmissions vs. time in RTB and GB-BTC under static channel conditions.

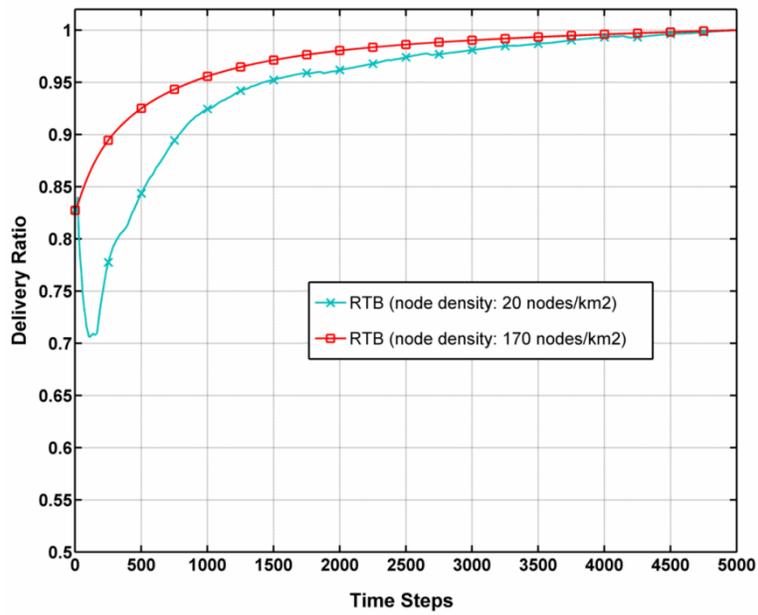

**Fig. 8.** The delivery ratio vs. time in RTB and Enhanced-RTB under static channel conditions.

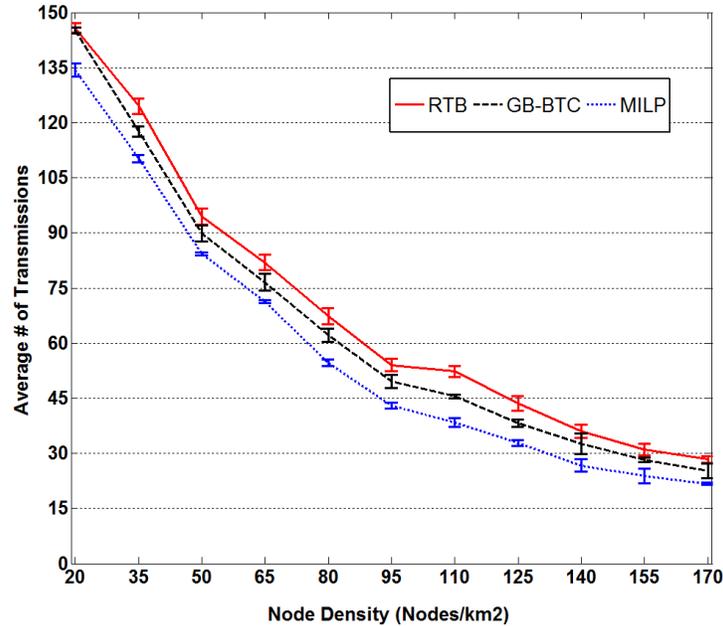

**Fig. 9.** The number of transmissions vs. various node densities under static channel conditions.

## 6. Conclusions and Outlook

The broadcast management problem in WANETs has been tackled with a wide assortment of heuristic-based methods reported in sundry publications. However, these methods lack a principled basis to explicitly factor the stochastic dynamics associated with variable link qualities into the problem formulation. In a departure from the prior art, this paper has presented a dynamic robust game formulation of this problem to capture the realistic effect posed by channel dynamics on message propagation. As an online solution to the decentralized stochastic control of broadcasting in a WANET with slow fading channels, a cognitive forwarding control mechanism has been proposed which is capable of inducing and maintaining strategic coordination between the forwarders' decisions. More specifically, we have deployed the strategic and non-stationary learning algorithm of *regret-tracking* which can converge to and track the correlated equilibria of the formulated game. An important aspect in our design has been to present a model-based variant of the learning algorithm which can exploit the available CSI for deriving a theoretically valid estimate of the transmission costs to speed up the learning process. As evidenced from the comparative numerical results, our proposed scheme can reduce the number of (re)transmissions and achieve a more balanced flow of messages while also maintaining near perfect delivery ratio in the presence of time-varying link qualities. As part of our plan for future work, we intend to extend our cognitive forwarding scheme to the other degradation categories for fading rapidity. In particular, we plan to come up with a stochastic game formulation of the broadcast management problem for WANETs with a fast fading regime of channel variations.